\documentclass[aps,prl,reprint,superscriptaddress,floatfix]{revtex4-2}
\usepackage[T1]{fontenc}
\usepackage{amsmath,amssymb,graphicx,bm,hyperref,booktabs}
\hypersetup{colorlinks=true,citecolor=blue,linkcolor=blue,urlcolor=blue,
pdftitle={Dihadron Production in Electron-Positron Annihilation at Next-to-Next-to-Leading Order in QCD},
pdfauthor={Long-Bin Chen; Wan-Li Ju; Hongxi Xing; Tong-Zhi Yang}}
\newcommand{\as}{a_s}

\newcommand{\MS}{\overline{\mathrm{MS}}}
\newcommand{\OS}{\mathrm{OS}}
\newcommand{\SameSign}{\mathrm{SS}}
\begin{document}
\title{Dihadron Production in Electron-Positron Annihilation at Next-to-Next-to-Leading Order in QCD}

\author{Long-Bin Chen}
\email{chenlb@gzhu.edu.cn}
\affiliation{School of Physics and Materials Science, Guangzhou University, Guangzhou 510006, China}

\author{Wan-Li Ju}
\email{wju@ualberta.ca}
\affiliation{Department of Physics, Hebei University, Baoding, 071002, China}
\affiliation{Hebei Key Laboratory of High-precision Computation and Application of Quantum Field Theory, Baoding, 071002, China}
\affiliation{Hebei Research Center of the Basic Discipline for Computational Physics, Baoding, 071002, China}
\affiliation{Department of Physics, University of Alberta, Edmonton AB T6G 2J1, Canada}

\author{Hongxi Xing}
\email{hxing@m.scnu.edu.cn}
\affiliation{State Key Laboratory of Nuclear Physics and Technology, Institute of Quantum Matter,
South China Normal University, Guangzhou 510006, China}
\affiliation{Guangdong Basic Research Center of Excellence for Structure and Fundamental Interactions of Matter,
Guangdong Provincial Key Laboratory of Nuclear Science, Guangzhou 510006, China}
\affiliation{Southern Center for Nuclear-Science Theory (SCNT), Institute of Modern Physics, Chinese Academy of Sciences, Huizhou 516000, China}

\author{Tong-Zhi Yang}
\email{tongzhi.yang@m.scnu.edu.cn}
\affiliation{State Key Laboratory of Nuclear Physics and Technology, Institute of Quantum Matter,
South China Normal University, Guangzhou 510006, China}
\affiliation{Guangdong Basic Research Center of Excellence for Structure and Fundamental Interactions of Matter,
Guangdong Provincial Key Laboratory of Nuclear Science, Guangzhou 510006, China}

\date{\today}
\begin{abstract}
Identified hadron pairs produced in electron-positron annihilation provide a unique channel to probe the flavor structure of hadronization inaccessible in single-hadron spectra. We present the first analytic calculation of the differential cross section in both hadron energy fractions at next-to-next-to-leading order (NNLO) in QCD. Incorporating fragmentation from both distinct two partons and a single parent parton, our calculation provides a stringent validation of collinear factorization and inhomogeneous evolution of dihadron fragmentation functions at NNLO. We demonstrate the crucial impact of NNLO corrections in comparison with Belle measurements of same-sign spectra and same-sign to opposite-sign ratios for $\pi\pi$, $\pi K$ and $KK$ pairs. These results enable fragmentation function analyses with dihadron data at NNLO accuracy.
\end{abstract}
\maketitle

\textit{Introduction.}---Fragmentation functions (FFs) are fundamental quantities in quantum chromodynamics (QCD) that describe how colored quarks and gluons hadronize into identified color singlet hadrons, serving as essential nonperturbative inputs into factorized high energy cross sections~\cite{Albino:2008gy,Metz:2016swz}. Precise determinations of FFs require high-order perturbative QCD calculations for hadron production in high energy collisions. For single inclusive hadron production in electron-positron annihilation (SIA), the perturbative coefficient functions are known at next-to-next-to-leading order (NNLO)~\cite{Rijken:1996ns,Mitov:2006wy} and next-to-next-to-next-to-leading order (N$^3$LO)~\cite{He:2025hin}. NNLO coefficients are likewise available for semi-inclusive deep inelastic scattering (SIDIS)~\cite{Bonino:2024qbh,Goyal:2023zdi,Goyal:2024emo,Dong:2026eas} and single inclusive hadron production in proton-proton collisions~\cite{Czakon:2025yti}. Combined with the NNLO timelike splitting functions~\cite{Mitov:2006ic,Moch:2007tx,Almasy:2011eq,Chen:2020uvt}, these theoretical advances have enabled high precision global determinations of FFs at NNLO~\cite{Anderle:2015lqa,Bertone:2017tyb,Soleymaninia:2020bsq,Borsa:2022vvp,AbdulKhalek:2022laj,Gao:2025hlm}.

Although SIA offers the cleanest environment for extracting FFs due to the absence of initial-state nonperturbative parton distributions, single-hadron spectra primarily constrain charge-weighted sums of quark and antiquark FFs. Separating individual quark flavors and distinguishing favored from unfavored fragmentation therefore remain challenging. Identifying a second hadron in opposite-sign, same-sign, or mixed-species pairs probes products of FFs and isolates distinct flavor and charge combinations. Moreover, this process provides direct access to dihadron fragmentation functions (DiFFs), extending the study of hadronization beyond the single-inclusive picture. Dihadron production in electron-positron collisions thus provides an indispensable probe of hadronization dynamics.

Precise measurements of hadron pair production by the Belle Collaboration~\cite{Belle:2015hut,Belle:2020pvy}, alongside ongoing programs at BESIII and planned facilities such as the FCC-$ee$~\cite{FCC:2025lpp}, CEPC~\cite{CEPCStudyGroup:2018ghi}, and STCF~\cite{Achasov:2023gey}, provide vital constraints on FFs beyond single-hadron data. Fully exploiting these datasets requires double-inclusive hard coefficients matching the NNLO accuracy of modern global FF analyses. Following the early study in Ref.~\cite{Altarelli:1979kv}, the angle integrated cross section differential in both hadron energies was computed at next-to-leading order (NLO) in Ref.~\cite{deFlorian:2003cg}. Since then, however, no NNLO calculation has been performed, due to the complexities of contributions both from two partons fragmenting independently and from a single parent parton fragmenting into both hadrons. The latter mechanism is encoded in DiFFs~\cite{Konishi:1979cb,Bianconi:1999cd}, whose scale evolution mixes inhomogeneously with products of single hadron FFs~\cite{Konishi:1979cb,Majumder:2004wh,Majumder:2004br,Ceccopieri:2007ip}. Analogous two-daughter kernels govern the evolution of track functions~\cite{Chang:2013rca,Li:2021zcf}, and their NLO expressions have been derived from multi-collinear splitting amplitudes~\cite{Chen:2022muj,Chen:2022pdu}. DiFFs differential in the pair invariant mass have been measured by Belle~\cite{Belle:2017rwm} and extracted in global analyses~\cite{Courtoy:2012ry,Pitonyak:2023gjx,Cocuzza:2023oam}, most recently at NNLO~\cite{Mahaut:2025hie}. While two-hadron angular correlations are known through $\mathcal{O}(\alpha_s^2)$ in the strong coupling $\alpha_s$ for opening angles strictly between $0$ and $\pi$~\cite{Ju:2025jxi}, integrating over the opening angle requires capturing all collinear singularities and the FF-DiFF mixing. At NNLO, this demands evaluating four-body phase-space integrals with two energy measurements across overlapping soft and collinear kinematic boundaries.

In this Letter, we present the first analytic NNLO calculation of the double energy coefficient functions for $e^+e^-$ dihadron production. In dimensional regularization ($d=4-2\epsilon$), all infrared and collinear poles from $1/\epsilon^4$ to $1/\epsilon$ cancel. This provides a stringent test of leading twist collinear factorization with FF-DiFF mixing at NNLO, and the single poles reproduce the NLO quark parent mixing kernels. We also identify a new charge odd contribution proportional to $e_qe_{q'}$, which is absent in SIA and directly probes cross flavor charge asymmetries of fragmentation. Our results enable double inclusive hadron production to the same perturbative accuracy as modern global FF fits. Comparing our predictions with Belle measurements of pion, kaon, and mixed pairs highlights the crucial role of NNLO corrections and demonstrates how absolute spectra and charge ratios place stringent constraints on flavor separated fragmentation dynamics.

\begin{figure}[t]
 \centering
 \begin{minipage}[c]{0.48\textwidth}
  \centering
  \makebox[0pt][r]{\raisebox{23.2mm}{\small(a)}\hspace{0.5mm}}%
  \includegraphics{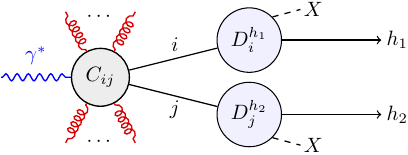}\par\bigskip
  \makebox[0pt][r]{\raisebox{21.3mm}{\small(b)}\hspace{0.5mm}}%
  \includegraphics{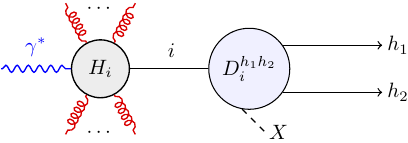}\par\bigskip
 \end{minipage}\hfill
 \caption{Schematic illustration of fragmentation mechanisms in Eq.~\eqref{eq:factorization}.
 (a) dihadron production from two distinct parent partons. (b) dihadron production from a single parent parton. Gray circles denote the hard coefficients $C_{ij}$ and $H_i$, blue circles denote the FFs and DiFFs with $X$ representing unobserved hadronic remnants, and curly lines indicate unobserved partonic radiation. 
}
 \label{fig:diagrams}
\end{figure}
\textit{Analytic calculations at NNLO.}---We consider dihadron ($h_1h_2$) production in electron-positron annihilation, $e^+e^-\to\gamma^*(q)\to h_1(P_1)h_2(P_2)+X$, where $X$ denotes the unobserved final state. At leading twist, the cross section differential in both hadron energies factorizes as~\cite{deFlorian:2003cg}
\begin{align}
 \frac{d^2\sigma}{\sigma_0\,dz_1dz_2}
 {}=&{}\sum_{ij}\int_{z_1}^{1}\!\frac{du}{u}\!\int_{z_2}^{1}\!\frac{dv}{v}
 C_{ij}(u,v)D_i^{h_1}\!\Big(\frac{z_1}{u}\Big)
 D_j^{h_2}\!\Big(\frac{z_2}{v}\Big)\notag\\
 &+\sum_i\int_{z_1+z_2}^{1}\!\frac{dw}{w^2}\,H_i(w)\,
 D_i^{h_1h_2}\!\Big(\frac{z_1}{w},\frac{z_2}{w}\Big).
 \label{eq:factorization}
\end{align}
Here $z_a=2q\cdot P_a/Q^2=2E_a/Q$ $(a=1,2)$ are the measured hadron energy fractions, with $E_a$ the hadron energies in the $e^+e^-$ center-of-mass frame and $Q=\sqrt{q^2}$ the center-of-mass energy. The normalization $\sigma_0=4\pi\alpha_{\rm em}^2N_c/(3Q^2)$ is the Born cross section for a single quark flavor of unit charge, where $\alpha_{\rm em}$ is the electromagnetic coupling and $N_c=3$ is the number of colors. The sums run over quarks, antiquarks, and gluons, with ordered hadron labels.

Eq.~\eqref{eq:factorization} consists of two contributions. The first term describes the independent single inclusive fragmentation of two well separated parent partons, as illustrated in Fig.~\ref{fig:diagrams}(a), where $D_i^h(z)$ denotes the FF of parton $i$ into hadron $h$, and $C_{ij}(u,v)$ is the hard coefficient for producing two partons $i$ and $j$ with energy fractions $u=2q\cdot p_i/Q^2$ and $v=2q\cdot p_j/Q^2$, respectively. The second term in Eq.~\eqref{eq:factorization} governs dihadron fragmentation, arising from two collimated hadrons produced by a single parent parton, as shown in Fig.~\ref{fig:diagrams}(b). Here $D_i^{h_1h_2}(z_1,z_2)$ is the collinear DiFF integrated over the relative transverse momentum of the hadron pair, and $H_i(w)$ is the hard coefficient for parent parton $i$ with energy fraction $w$. This coefficient coincides with that of SIA~\cite{Rijken:1996ns,Mitov:2006wy} and is also connected to the energy-energy correlator (EEC)~\cite{Basham:1978bw} in the collinear limit~\cite{Lee:2025okn,Kang:2025zto}. Through $\mathcal{O}(\alpha_s^2)$, it takes the form $H_i=\sum_q e_q^2\,H_i^{[q]}$.

\begin{figure}[t]
 \centering
  \includegraphics{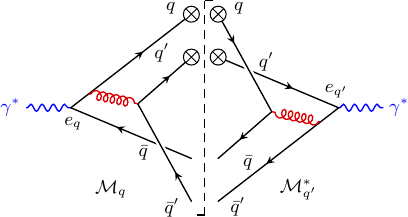}%
 \caption{Representative cut diagram for the new $e_qe_{q'}$ interference contribution to $C_{ij}$ at $\mathcal{O}(\alpha_s^2)$ with $\otimes$ marking the observed partons $q$ and $q'$. 
}
 \label{fig:feyndiagrams}
\end{figure}

Organized by the electric charges of the primary quarks coupled to the electromagnetic current, the hard coefficients $C_{ij}$ decompose as
\begin{align}
C_{ij}=\sum_q e_q^2\,C_{ij}^{[q]}+\frac{\as^2}{2}\sum_{q\neq q'}e_qe_{q'}\,I_{ij}^{[qq']},
\end{align}
where $\as=\alpha_s/(2\pi)$ denotes the strong coupling. The interference term $I_{ij}^{[qq']}=I_{ij}^{[q'q]}$, obtained by integrating $2\,\mathrm{Re}(\mathcal M_q\mathcal M_{q'}^*)$ at fixed $u$ and $v$, describes the case in which the photon couples to different quark lines in the amplitude and its complex conjugate, as illustrated in Fig.~\ref{fig:feyndiagrams}. This new contribution first appears at $\mathcal{O}(\alpha_s^2)$ and is free of infrared poles. Reversing the orientation of either quark line flips the sign of $I_{ij}^{[qq']}$, so this term vanishes in SIA but contributes to Eq.~\eqref{eq:factorization} through the charge-odd products $\Delta_q^{h_1}\Delta_{q'}^{h_2}$ with $\Delta_q^h=D_q^h-D_{\bar q}^h$.

We expand each hard coefficient $F\in\{C_{ij},H_i\}$ as $F=F^{(0)}+\as F^{(1)}+\as^2F^{(2)}+\mathcal{O}(\as^3)$, working in $d=4-2\epsilon$ dimensions in the modified minimal subtraction ($\MS$) scheme. At NNLO, $F^{(2)}$ receives double-real (RR), real-virtual (RV), and double-virtual (VV) contributions. We extract the VV contributions from the two-loop quark form factor~\cite{Gehrmann:2005pd} and the squared one-loop amplitude. To evaluate the RR and RV corrections, we generate the Feynman diagrams and perform the amplitude algebra with QGRAF~\cite{Nogueira:1991ex}, FeynArts~\cite{Hahn:2000kx}, FORM~\cite{Ruijl:2017dtg}, and FeynCalc~\cite{Shtabovenko:2023idz}. Reverse unitarity~\cite{Anastasiou:2002yz} is employed to represent the on-shell conditions and energy measurements as cut propagators, followed by integration-by-parts reduction~\cite{Chetyrkin:1981qh,Laporta:2000dsw} with Kira~\cite{Maierhofer:2017gsa,Lange:2025fba}, yielding 18 and 7 master integrals in the RR and RV sectors, respectively. Their canonical differential equations~\cite{Gehrmann:1999as,Henn:2013pwa} exhibit the linear alphabet
\begin{equation}
\{u,v,1-u,1-v,1-u-v,u+v,1+u,1+v,2-u-v\}.
 \label{eq:alphabet}
\end{equation}
Boundary constants are fixed by soft limits and the regularity of the phase-space integrals. The resulting solutions in both kinematic regions $u+v\gtrless 1$ are expressed in terms of generalized polylogarithms (GPLs) with the physical timelike analytic continuation~\cite{Goncharov:2001iea,Vollinga:2004sn}, and have been numerically cross-checked against AMFlow~\cite{Liu:2022chg,Huang:2026rjb} at randomly sampled $(u,v)$ points.

The renormalization of the hard coefficients proceeds in two steps. The first step consists of the UV renormalization of the strong coupling constant, i.e., $ \alpha_{s}^{b}\mu_{0}^{2\epsilon} = Z_{\alpha_{s}}\mu_R^{2\epsilon}\alpha_{s}(\mu_R)$, 
where $Z_{\alpha_{s}}$ denotes the counterterm in the $\overline{\mathrm{MS}}$ scheme. In the second step, the residual infrared-collinear (IRC) divergences in the hard coefficients are absorbed by the renormalization of  the single-hadron and dihadron fragmentation functions~\cite{deFlorian:2003cg,Majumder:2004wh,Ceccopieri:2007ip}, 
\begin{align}
\label{eq:def:Hbare:ren}
 H_i (w)= &\int_{w}^{1} \frac{\mathrm{d}t}{t}\, H_{a}^{B}(t)\, Z^{P}_{a/ i}\!\left(\frac{w}{t}\right), \\
  \label{eq:def:Cbare:ren}
 C_{ij}(u,v) = &\int_{u}^{1} \frac{\mathrm{d}t_1}{t_1} \int_{v}^{1} \frac{\mathrm{d}t_2}{t_2}\, C_{ab}^{B}(t_1,t_2)  Z^{D}_{a/ i}\!\left(\frac{u}{t_1}\right)Z^{D}_{ b/ j}\!\left(\frac{v}{t_2}\right)\nonumber\\
  &
   +
\int_{u+v}^{1} \frac{\mathrm{d}t}{t^2}\, H_{a}^{B}(t)\, Z^{K}_{a/ ij}\!\left(\frac{u}{t}, \frac{v}{t}\right),
\end{align}
where the summation is implied over repeated parton indices $a$ and $b$ for brevity. Herein, wide angle soft divergences together with those arising from the soft-collinear overlap region cancel out in the inclusive sum over cuts. As a result, the bare hard coefficients $C^B$ and $H^B$ contain at most $1/\epsilon^2$ poles through $\mathcal{O}(\alpha_s^2)$ originating from collinear splittings. To absorb these poles, the renormalization factors $Z^P_{a/i}$ and $Z^D_{a/i}$ are introduced to account for collinear splittings associated with single-parton fragmentation.
 Both $Z^P_{a/i}$ and $Z^D_{a/i}$ are governed by the single-parton time-like splitting functions~\cite{Curci:1980uw,Furmanski:1980cm,Floratos:1981hs,Stratmann:1996hn}, 
 while only  differing in the momentum fraction arguments in their convolution integrals. By contrast, $Z^{K}_{a/ ij}$ addresses a distinct kinematic configuration in which two collimated hadrons originate from a single parent parton.  
 The expression of $Z^{K}_{a/ ij}$ is available through $\mathcal{O}(\alpha_s^2)$~\cite{deFlorian:2003cg,Chen:2022pdu}  till now.
 
With these renormalization factors, the single hadron FFs and DiFFs renormalize in a coupled manner on the basis of $(D_a^{h_1h_2},D_a^{h_1}D_b^{h_2})$. 
The renormalized DiFFs are  thus subject to the evolution equation~\cite{Konishi:1979cb,Majumder:2004wh,Majumder:2004br,Ceccopieri:2007ip,Chen:2022pdu},
\begin{align}
 \frac{dD_i^{h_1h_2}(z_1,z_2)}{d\ln\mu_F^2}
 =&\int_{z_1+z_2}^{1}\!\frac{dw}{w^2}\,P_{i/ j}(w)\,
 D_j^{h_1h_2}\!\Big(\frac{z_1}{w},\frac{z_2}{w}\Big)\notag\\
 &+\int_{z_1}^{1-z_2}\!\frac{dx}{x}
 \int_{z_2}^{1-x}\!\frac{dy}{y}\,K^{i}_{jk}(x,y) \nonumber\\
 &\times D_j^{h_1}\!\Big(\frac{z_1}{x}\Big)
 D_k^{h_2}\!\Big(\frac{z_2}{y}\Big),
 \label{eq:DiFFevolution}
\end{align}
where  $\mu_F$ stands for the factorization scale and is suppressed in the fragmentation functions above for brevity.
 $P_{i/ j}$ and $K^{i}_{jk}$  are splitting kernels induced by the renormalization counterterms $Z^P_{i/ j}$ and $Z^{K}_{i/jk}$, respectively. 

We compare the pole structure in our UV-renormalized hard coefficients $H^B$ and $C^B$ against those predicted by the single-parton time-like splitting functions~\cite{Curci:1980uw,Furmanski:1980cm,Floratos:1981hs,Stratmann:1996hn} and multiparton splitting kernels in Refs.~\cite{deFlorian:2003cg,Chen:2022pdu}, finding exact analytic agreement. This agreement not only supports the correctness of our calculation, but also provides a non-trivial validation of leading-twist factorization in Eq.~\eqref{eq:factorization}. The second check relies on the momentum sum rules for FFs and DiFFs, which, together with energy-momentum conservation, relate the two renormalized hard coefficients via
\begin{equation}
  \sum_j \int_0^1 \mathrm{d}v\, v\, C_{ij}^{(2)}(u,v) = (2-u) H_i^{(2)}(u),
  \label{eq:SIAprojection}
\end{equation}
where the sum runs over all parton flavors $j$. Because partonic energy fractions are normalized to $Q/2$, they sum to $2$, leaving $2-u$ for the remaining partons once the first is fixed at $u$. Taking $H_i^{(2)}(u)$ from Refs.~\cite{Rijken:1996ns,Mitov:2006wy} and evaluating the integrals analytically with \texttt{PolyLogTools}~\cite{Duhr:2019tlz}, we verify Eq.~\eqref{eq:SIAprojection} across the non-singlet, pure-singlet, and gluon channels, including all endpoint distributions.

The verified two variable parts of $C_{ij}^{(2)}$ contain only logarithms and dilogarithms, i.e., functions of weight at most two in the letters of Eq.~\eqref{eq:alphabet}. Weight three functions, $\mathrm{Li}_3$ and $\zeta_3$, appear only in the coefficients of distributions supported on $u=1, v=1$, and $u+v=1$. The $e_qe_{q'}$ interference contains only rational functions and single logarithms. The ancillary files accompanying this paper provide the analytic expressions of $C_{ij}$ and $H_i$, including our GPL and distribution conventions.

\begin{figure*}[t]
 \centering
 \includegraphics[width=\textwidth]{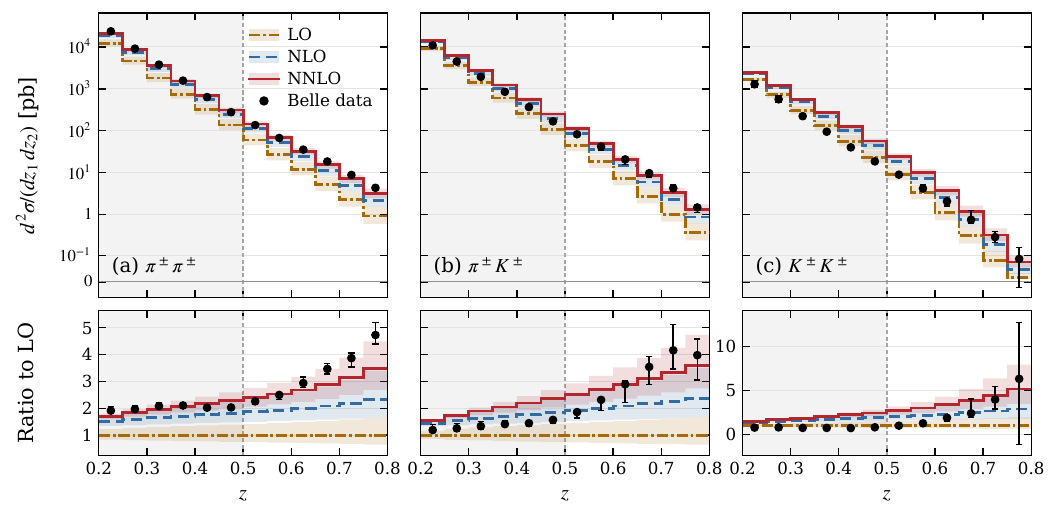}
 \caption{Same-sign spectra (upper panels) and ratios to LO (lower panels)
 for (a) $\pi\pi$, (b) $\pi K$, and (c) $KK$ pairs at $Q=10.58$ GeV. Here $z$ labels the square-bin center. The theory calculations use NPC23
 NNLO FFs at all perturbative orders, and the DiFFs are set to zero. The bands represent scale uncertainties from $\mu_R=\mu_F\in[Q/2,2Q]$, with central curves at $Q$.
 Solid circles show Belle's weak decay subtracted data~\cite{Belle:2020pvy}. In the lower panels, predictions at each perturbative order and Belle data are divided by the LO prediction at $Q$.}
 \label{fig:absolute}
\end{figure*}

\textit{Phenomenological study.}---To assess the phenomenological impact of our results, we perform numerical calculations at NNLO and compare with Belle's weak decay subtracted spectra at $Q=10.58~\text{GeV}$~\cite{Belle:2020pvy}, without hemisphere or thrust cuts. Because NNLO DiFFs for general hadron pairs remain undetermined, we compute predictions from the single hadron fragmentation channel alone (the first term in Eq.~\eqref{eq:factorization}). The calculation uses five massless quark flavors with the primary bottom-quark coupling to the photon switched off, and the hadron mass corrections are neglected. We employ NNLO FFs at all perturbative orders, so that the $K$-factors isolate the higher-order corrections to the hard coefficients. For $\pi \pi$ and $K K$, we denote the charge-averaged same-sign (SS) and opposite-sign (OS) cross sections by $d\sigma_{\SameSign}=\tfrac12(d\sigma_{h_1^+h_2^+}+d\sigma_{h_1^-h_2^-})$ and $d\sigma_{\OS}=\tfrac12(d\sigma_{h_1^+h_2^-}+d\sigma_{h_1^-h_2^+})$, with ordered hadron labels and no additional identical particle factor. For $\pi K$, we label the pion first and replace $1/2$ by $1$ in both expressions.

In Fig.~\ref{fig:absolute}, we compare our predictions with Belle's absolute same-sign dihadron yields for $\pi\pi$, $\pi K$, and $KK$ in diagonal square bins ($z_1=z_2=z$) using NPC23 FFs~\cite{Gao:2025hlm}. The bands indicate scale uncertainties obtained by varying the renormalization and factorization scales simultaneously by a factor of $2$ ($\mu_R=\mu_F\in[Q/2,2Q]$). The results clearly show relative scale uncertainties decrease from NLO to NNLO, although the residual scale dependence grows with $z$. For $z>0.5$, where FF-DiFF mixing does not contribute, this growth reflects enhanced logarithms near the partonic threshold $u,v\to1$, whereas for $z<0.5$ the factorization scale logarithms arising from FF-DiFF mixing in $C_{ij}$ remain uncompensated until DiFFs are included.

To quantify the impact of higher order corrections, we show in the lower panels of Fig.~\ref{fig:absolute} the $K$-factors, defined as the ratios of the cross sections at each perturbative order to the central LO prediction. These $K$-factors reveal increasingly large contributions from higher-order terms. The NNLO enhancement over NLO grows with $z$ from $13\%$ to $50\%$ for $\pi\pi$, $8\%$ to $52\%$ for $\pi K$, and $4\%$ to $79\%$ for $KK$. Such substantial corrections are essential for a meaningful comparison with the Belle data. The region $z>0.5$, where the DiFF contribution vanishes identically, provides a direct comparison between NNLO prediction and experiment. The agreement for $\pi\pi$ within scale uncertainties up to $z\approx0.7$ reflects how well pion FFs are constrained, whereas the discrepancy for $KK$ highlights the constraining power of kaon pair production on kaon FFs. In the shaded region $z<0.5$, the agreement for $\pi\pi$ indicates a relatively modest DiFF contribution compared with single hadron fragmentation, while the excess above the $\pi K$ and $KK$ data, already visible for $KK$ at $z>0.5$, calls for a simultaneous determination of kaon FFs and DiFFs.

\begin{figure*}[t]
 \centering
 \includegraphics[width=\textwidth]{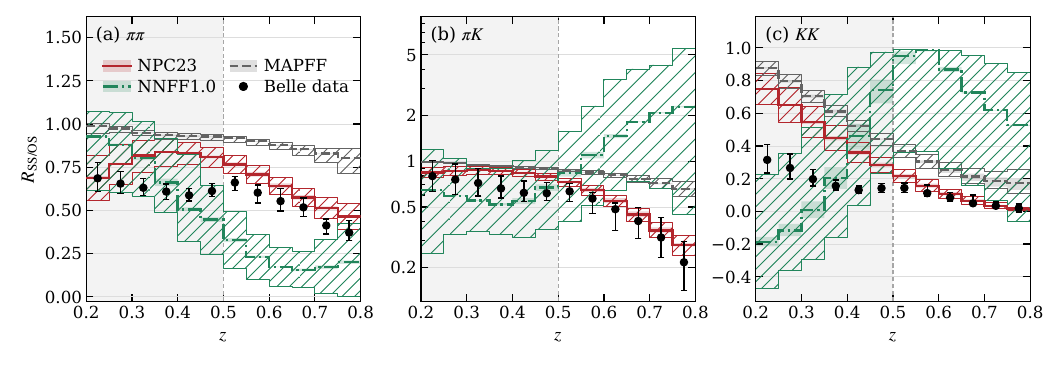}
 \caption{NNLO same-sign to opposite-sign ratios $R_{\SameSign/\OS}$ for (a) $\pi\pi$, (b) $\pi K$, and (c) $KK$. Curves show central predictions at $\mu=Q$, while filled bands represent scale variations $\mu_R=\mu_F\in[Q/2,2Q]$. Hatched bands denote FF uncertainties for NPC23 (red), NNFF1.0 (green), and MAPFF (gray). 
 }
 \label{fig:ratio}
\end{figure*}

To probe the discriminating power of dihadron production on FFs, we compute the same-sign to opposite-sign charge ratio at NNLO,
\begin{equation}
 R_{\SameSign/\OS}(B)=\frac{\int_B d\sigma_\SameSign}{\int_B d\sigma_\OS},
 \label{eq:ratio}
\end{equation}
evaluated over each kinematic bin $B$, with $d\sigma_{\SameSign}$ and $d\sigma_{\OS}$ defined above. The same scales and FF members are used in the numerator and denominator, with scale and FF uncertainties evaluated separately. Unlike SIA spectra, which are sensitive only to charge-summed combinations ($\Sigma_q^h=D_q^h+D_{\bar q}^h$), this ratio also probes the charge asymmetric combinations $\Delta_q^h$, providing unique flavor sensitivity in $e^+e^-$ collisions. Beyond the same flavor products $\Delta_q^{h_1}\Delta_q^{h_2}$ present from LO, the $e_qe_{q'}$ interference supplies the only cross flavor products $\Delta_q^{h_1}\Delta_{q'}^{h_2}$ at this order. We compare predictions using NPC23~\cite{Gao:2025hlm}, MAPFF~\cite{AbdulKhalek:2022laj}, and NNFF1.0~\cite{Bertone:2017tyb}, all of which successfully describe SIA data. As shown by the filled bands in Fig.~\ref{fig:ratio}, scale uncertainties cancel substantially in the ratio across all three FF sets. The hatched bands represent the FF uncertainties, reflecting the distinct correlated and anticorrelated flavor structures of the respective fits.

In the $z>0.5$ region, both NPC23 and MAPFF reproduce the downward trend of all three Belle ratios, with NPC23 lying closest to the data, whereas NNFF1.0 exhibits the largest discrepancy. This difference stems from the underlying datasets, while NPC23 and MAPFF incorporate charge identified SIDIS measurements, NNFF1.0 fits only charge summed SIA data, so its charge separated grids rely on additional assumptions~\cite{Bertone:2017tyb}. In particular, NNFF1.0 and NPC23 differ in how $\Delta_q^h/\Sigma_q^h$ behaves for pions and kaons, so their $\pi K$ ratios evolve with $z$ in opposite directions. In the $z<0.5$ region, DiFF contributions do not generally cancel in the ratio. With DiFFs omitted, NPC23 overestimates the $KK$ ratio by roughly a factor of $2$, and the other two FF sets likewise fail to describe the data, indicating the necessity of including DiFFs to match the precision of existing measurements.

\textit{Conclusion.}---We have presented the first analytic NNLO calculation in QCD of the angle integrated dihadron production cross section differential in both hadron energy fractions in electron-positron annihilation. Incorporating both two parton fragmentation and single parton dihadron fragmentation, our results achieve the complete cancellation of infrared and collinear singularities from $1/\epsilon^4$ to $1/\epsilon$ and provide a nontrivial validation of leading twist collinear factorization with inhomogeneous FF-DiFF mixing at $\mathcal{O}(\alpha_s^2)$. Our calculation is validated by numerical checks of the master integrals, the known splitting kernels $P$ and $K$, and the momentum sum rule in Eq.~\eqref{eq:SIAprojection}. We have further identified a finite, charge odd $e_qe_{q'}$ interference contribution, absent in SIA, which provides direct sensitivity to cross flavor charge asymmetries of fragmentation.

Phenomenological comparisons with Belle measurements at $Q=10.58~\text{GeV}$~\cite{Belle:2020pvy} demonstrate the crucial impact of NNLO corrections. Relative to NLO, the NNLO contributions reduce perturbative scale uncertainties and enhance the same-sign yields at large $z$ by up to $50\%$ for $\pi\pi$, $52\%$ for $\pi K$, and $79\%$ for $KK$, which are essential for the comparison with the data. In the region $z_1+z_2>1$, where DiFFs vanish by kinematics, the charge ratio $R_{\SameSign/\OS}$ exhibits minimal scale dependence and clearly discriminates among modern FF sets, highlighting the sensitivity of dihadron observables to charge-asymmetric and strange-quark fragmentation. In the region $z_1+z_2\le 1$, discrepancies in the kaon channels when DiFFs are omitted demonstrate the need to determine FFs and DiFFs jointly. By elevating double-inclusive hadron production to the same NNLO accuracy as single inclusive global fits, our results provide a new opportunity to incorporating dihadron spectra and charge ratios, including off-diagonal bins and kaon channels from Belle as well as future high precision programs at BESIII, STCF, CEPC, and FCC-$ee$, into next generation global analyses for full flavor- and charge-separated determinations of FFs and DiFFs.

\begin{acknowledgments}
We thank Xiaohui Liu and Hua-Xing Zhu for useful discussions. The work of H.X. was supported by National Natural
Science Foundation of China with Projects No. 12525508 and No. 12475139. The work of L.B.C was supported by the National Natural Science Foundation of China under the Grants No. 12622504, and by the Guangdong Basic and Applied Basic Research Foundation under grant No. 2025B1515020009. The work of T.-Z.Y. was supported by the start-up funding of South China Normal University under Grant No. 43102674.
\end{acknowledgments}


\begin{thebibliography}{10}

\bibitem{Albino:2008gy}
S.~Albino,
\newblock Rev. Mod. Phys. {\bf 82}, 2489 (2010), arXiv:0810.4255.

\bibitem{Metz:2016swz}
A.~Metz and A.~Vossen,
\newblock Prog. Part. Nucl. Phys. {\bf 91}, 136 (2016), arXiv:1607.02521.

\bibitem{Rijken:1996ns}
P.~J. Rijken and W.~L. van Neerven,
\newblock Nucl. Phys. B {\bf 487}, 233 (1997), arXiv:hep-ph/9609377.

\bibitem{Mitov:2006wy}
A.~Mitov and S.-O. Moch,
\newblock Nucl. Phys. B {\bf 751}, 18 (2006), arXiv:hep-ph/0604160.

\bibitem{He:2025hin}
C.-Q. He, H.~Xing, T.-Z. Yang, and H.~X. Zhu,
\newblock Phys. Rev. Lett. {\bf 135}, 101901 (2025), arXiv:2503.20441.

\bibitem{Bonino:2024qbh}
L.~Bonino, T.~Gehrmann, and G.~Stagnitto,
\newblock Phys. Rev. Lett. {\bf 132}, 251901 (2024), arXiv:2401.16281.

\bibitem{Goyal:2023zdi}
S.~Goyal, S.-O. Moch, V.~Pathak, N.~Rana, and V.~Ravindran,
\newblock Phys. Rev. Lett. {\bf 132}, 251902 (2024), arXiv:2312.17711.

\bibitem{Goyal:2024emo}
S.~Goyal {\em et~al.},
\newblock Phys. Rev. D {\bf 111}, 094007 (2025), arXiv:2412.19309.

\bibitem{Dong:2026eas}
L.~Dong {\em et~al.},
\newblock Phys. Rev. Lett. {\bf 137}, 081902 (2026), arXiv:2602.22972.

\bibitem{Czakon:2025yti}
M.~Czakon, T.~Generet, A.~Mitov, and R.~Poncelet,
\newblock Phys. Rev. Lett. {\bf 135}, 17 (2025), arXiv:2503.11489.

\bibitem{Mitov:2006ic}
A.~Mitov, S.~Moch, and A.~Vogt,
\newblock Phys. Lett. B {\bf 638}, 61 (2006), arXiv:hep-ph/0604053.

\bibitem{Moch:2007tx}
S.~Moch and A.~Vogt,
\newblock Phys. Lett. B {\bf 659}, 290 (2008), arXiv:0709.3899.

\bibitem{Almasy:2011eq}
A.~A. Almasy, S.~Moch, and A.~Vogt,
\newblock Nucl. Phys. B {\bf 854}, 133 (2012), arXiv:1107.2263.

\bibitem{Chen:2020uvt}
H.~Chen, T.-Z. Yang, H.~X. Zhu, and Y.~J. Zhu,
\newblock Chin. Phys. C {\bf 45}, 043101 (2021), arXiv:2006.10534.

\bibitem{Anderle:2015lqa}
D.~P. Anderle, F.~Ringer, and M.~Stratmann,
\newblock Phys. Rev. D {\bf 92}, 114017 (2015), arXiv:1510.05845.

\bibitem{Bertone:2017tyb}
NNPDF, V.~Bertone, S.~Carrazza, N.~P. Hartland, E.~R. Nocera, and J.~Rojo,
\newblock Eur. Phys. J. C {\bf 77}, 516 (2017), arXiv:1706.07049.

\bibitem{Soleymaninia:2020bsq}
M.~Soleymaninia, M.~Goharipour, H.~Khanpour, and H.~Spiesberger,
\newblock Phys. Rev. D {\bf 103}, 054045 (2021), arXiv:2008.05342.

\bibitem{Borsa:2022vvp}
I.~Borsa, R.~Sassot, D.~de~Florian, M.~Stratmann, and W.~Vogelsang,
\newblock Phys. Rev. Lett. {\bf 129}, 012002 (2022), arXiv:2202.05060.

\bibitem{AbdulKhalek:2022laj}
MAP (Multi-dimensional Analyses of Partonic distributions), R.~Abdul~Khalek,
  V.~Bertone, A.~Khoudli, and E.~R. Nocera,
\newblock Phys. Lett. B {\bf 834}, 137456 (2022), arXiv:2204.10331.

\bibitem{Gao:2025hlm}
J.~Gao, X.~Shen, H.~Xing, Y.~Zhao, and B.~Zhou,
\newblock Phys. Rev. Lett. {\bf 135}, 041902 (2025), arXiv:2502.17837.

\bibitem{Belle:2015hut}
Belle, R.~Seidl {\em et~al.},
\newblock Phys. Rev. D {\bf 92}, 092007 (2015), arXiv:1509.00563.

\bibitem{Belle:2020pvy}
Belle, R.~Seidl {\em et~al.},
\newblock Phys. Rev. D {\bf 101}, 092004 (2020), arXiv:2001.10194.

\bibitem{FCC:2025lpp}
FCC, M.~Benedikt {\em et~al.},
\newblock Eur. Phys. J. C {\bf 85}, 1468 (2025), arXiv:2505.00272,
\newblock [Erratum: Eur.Phys.J.C 86, 844 (2026)].

\bibitem{CEPCStudyGroup:2018ghi}
CEPC Study Group, M.~Dong {\em et~al.},
\newblock IHEP Report No. IHEP-CEPC-DR-2018-02, 2018 (unpublished),
  arXiv:1811.10545.

\bibitem{Achasov:2023gey}
M.~Achasov {\em et~al.},
\newblock Front. Phys. (Beijing) {\bf 19}, 14701 (2024), arXiv:2303.15790.

\bibitem{Altarelli:1979kv}
G.~Altarelli, R.~K. Ellis, G.~Martinelli, and S.-Y. Pi,
\newblock Nucl. Phys. B {\bf 160}, 301 (1979).

\bibitem{deFlorian:2003cg}
D.~de~Florian and L.~Vanni,
\newblock Phys. Lett. B {\bf 578}, 139 (2004), arXiv:hep-ph/0310196.

\bibitem{Konishi:1979cb}
K.~Konishi, A.~Ukawa, and G.~Veneziano,
\newblock Nucl. Phys. B {\bf 157}, 45 (1979).

\bibitem{Bianconi:1999cd}
A.~Bianconi, S.~Boffi, R.~Jakob, and M.~Radici,
\newblock Phys. Rev. D {\bf 62}, 034008 (2000), arXiv:hep-ph/9907475.

\bibitem{Majumder:2004wh}
A.~Majumder and X.-N. Wang,
\newblock Phys. Rev. D {\bf 70}, 014007 (2004), arXiv:hep-ph/0402245.

\bibitem{Majumder:2004br}
A.~Majumder and X.-N. Wang,
\newblock Phys. Rev. D {\bf 72}, 034007 (2005), arXiv:hep-ph/0411174.

\bibitem{Ceccopieri:2007ip}
F.~A. Ceccopieri, M.~Radici, and A.~Bacchetta,
\newblock Phys. Lett. B {\bf 650}, 81 (2007), arXiv:hep-ph/0703265.

\bibitem{Chang:2013rca}
H.-M. Chang, M.~Procura, J.~Thaler, and W.~J. Waalewijn,
\newblock Phys. Rev. Lett. {\bf 111}, 102002 (2013), arXiv:1303.6637.

\bibitem{Li:2021zcf}
Y.~Li, I.~Moult, S.~S. van Velzen, W.~J. Waalewijn, and H.~X. Zhu,
\newblock Phys. Rev. Lett. {\bf 128}, 182001 (2022), arXiv:2108.01674.

\bibitem{Chen:2022muj}
H.~Chen {\em et~al.},
\newblock Phys. Rev. D {\bf 111}, 076021 (2025), arXiv:2210.10061.

\bibitem{Chen:2022pdu}
H.~Chen {\em et~al.},
\newblock JHEP {\bf 2023}, 185 (2023), arXiv:2210.10058.

\bibitem{Belle:2017rwm}
Belle, R.~Seidl {\em et~al.},
\newblock Phys. Rev. D {\bf 96}, 032005 (2017), arXiv:1706.08348.

\bibitem{Courtoy:2012ry}
A.~Courtoy, A.~Bacchetta, M.~Radici, and A.~Bianconi,
\newblock Phys. Rev. D {\bf 85}, 114023 (2012), arXiv:1202.0323.

\bibitem{Pitonyak:2023gjx}
D.~Pitonyak, C.~Cocuzza, A.~Metz, A.~Prokudin, and N.~Sato,
\newblock Phys. Rev. Lett. {\bf 132}, 011902 (2024), arXiv:2305.11995.

\bibitem{Cocuzza:2023oam}
JAM, C.~Cocuzza {\em et~al.},
\newblock Phys. Rev. Lett. {\bf 132}, 091901 (2024), arXiv:2306.12998.

\bibitem{Mahaut:2025hie}
MAP (Multi-dimensional Analyses of Partonic distributions), V.~Mahaut {\em
  et~al.},
\newblock JHEP {\bf 2026}, 051 (2026), arXiv:2509.11855.

\bibitem{Ju:2025jxi}
W.-L. Ju, Z.-Y. Shu, T.-Z. Yang, Z.-H. Zhang, and H.~X. Zhu,
\newblock JHEP {\bf 2025}, 051 (2025), arXiv:2506.11463.

\bibitem{Basham:1978bw}
C.~L. Basham, L.~S. Brown, S.~D. Ellis, and S.~T. Love,
\newblock Phys. Rev. Lett. {\bf 41}, 1585 (1978).

\bibitem{Lee:2025okn}
K.~Lee and I.~W. Stewart,
\newblock Phys. Rev. Lett. {\bf 136}, 081902 (2026), arXiv:2507.11495.

\bibitem{Kang:2025zto}
Z.-B. Kang, A.~Metz, D.~Pitonyak, and C.~Zhang,
\newblock Phys. Rev. Lett. {\bf 136}, 081905 (2026), arXiv:2507.17444.

\bibitem{Gehrmann:2005pd}
T.~Gehrmann, T.~Huber, and D.~Maitre,
\newblock Phys. Lett. B {\bf 622}, 295 (2005), arXiv:hep-ph/0507061.

\bibitem{Nogueira:1991ex}
P.~Nogueira,
\newblock J. Comput. Phys. {\bf 105}, 279 (1993).

\bibitem{Hahn:2000kx}
T.~Hahn,
\newblock Comput. Phys. Commun. {\bf 140}, 418 (2001), arXiv:hep-ph/0012260.

\bibitem{Ruijl:2017dtg}
B.~Ruijl, T.~Ueda, and J.~Vermaseren,
\newblock {FORM version 4.2}, 2017, arXiv:1707.06453.

\bibitem{Shtabovenko:2023idz}
V.~Shtabovenko, R.~Mertig, and F.~Orellana,
\newblock Comput. Phys. Commun. {\bf 306}, 109357 (2025), arXiv:2312.14089.

\bibitem{Anastasiou:2002yz}
C.~Anastasiou and K.~Melnikov,
\newblock Nucl. Phys. B {\bf 646}, 220 (2002), arXiv:hep-ph/0207004.

\bibitem{Chetyrkin:1981qh}
K.~G. Chetyrkin and F.~V. Tkachov,
\newblock Nucl. Phys. B {\bf 192}, 159 (1981).

\bibitem{Laporta:2000dsw}
S.~Laporta,
\newblock Int. J. Mod. Phys. A {\bf 15}, 5087 (2000), arXiv:hep-ph/0102033.

\bibitem{Maierhofer:2017gsa}
P.~Maierh{\"o}fer, J.~Usovitsch, and P.~Uwer,
\newblock Comput. Phys. Commun. {\bf 230}, 99 (2018), arXiv:1705.05610.

\bibitem{Lange:2025fba}
F.~Lange, J.~Usovitsch, and Z.~Wu,
\newblock Comput. Phys. Commun. {\bf 322}, 109999 (2026), arXiv:2505.20197.

\bibitem{Gehrmann:1999as}
T.~Gehrmann and E.~Remiddi,
\newblock Nucl. Phys. B {\bf 580}, 485 (2000), arXiv:hep-ph/9912329.

\bibitem{Henn:2013pwa}
J.~M. Henn,
\newblock Phys. Rev. Lett. {\bf 110}, 251601 (2013), arXiv:1304.1806.

\bibitem{Goncharov:2001iea}
A.~B. Goncharov,
\newblock {Multiple polylogarithms and mixed Tate motives}, 2001,
  arXiv:math/0103059.

\bibitem{Vollinga:2004sn}
J.~Vollinga and S.~Weinzierl,
\newblock Comput. Phys. Commun. {\bf 167}, 177 (2005), arXiv:hep-ph/0410259.

\bibitem{Liu:2022chg}
X.~Liu and Y.-Q. Ma,
\newblock Comput. Phys. Commun. {\bf 283}, 108565 (2023), arXiv:2201.11669.

\bibitem{Huang:2026rjb}
R.-J. Huang, X.~Liu, and Y.-Q. Ma,
\newblock {AMFlow 2.0: significant algorithmic and software improvements for
  Feynman integral evaluation}, 2026, arXiv:2607.08477.

\bibitem{Curci:1980uw}
G.~Curci, W.~Furmanski, and R.~Petronzio,
\newblock Nucl. Phys. B {\bf 175}, 27 (1980).

\bibitem{Furmanski:1980cm}
W.~Furmanski and R.~Petronzio,
\newblock Phys. Lett. B {\bf 97}, 437 (1980).

\bibitem{Floratos:1981hs}
E.~G. Floratos, C.~Kounnas, and R.~Lacaze,
\newblock Nucl. Phys. B {\bf 192}, 417 (1981).

\bibitem{Stratmann:1996hn}
M.~Stratmann and W.~Vogelsang,
\newblock Nucl. Phys. B {\bf 496}, 41 (1997), arXiv:hep-ph/9612250.

\bibitem{Duhr:2019tlz}
C.~Duhr and F.~Dulat,
\newblock JHEP {\bf 2019}, 135 (2019), arXiv:1904.07279.

\end{thebibliography}

\end{document}